\documentclass[12pt, draftclsnofoot, onecolumn]{IEEEtran}

\usepackage{epsfig}
\usepackage{epstopdf}
\usepackage{times}
\usepackage{bm}
\usepackage{amsmath,amssymb} 
\usepackage{subfigure}
\usepackage{algorithmic}
\usepackage{algorithm}
\usepackage{caption2}
\usepackage{textcomp}
\usepackage{graphicx}

\usepackage[table,xcdraw]{xcolor}

\usepackage{amsmath}
\usepackage{authblk}

\begin{document}

\title{Optimal Resource Allocation in Multicast Device-to-Device Communications Underlaying LTE Networks }

\author[1]{Hadi Meshgi}
\author[1]{Dongmei Zhao}
\author[2]{Rong Zheng}
\affil[1]{Department of Electrical and Computer Engineering, McMaster University}
\affil[2]{Department of Computing and Software, McMaster University}

\renewcommand\Authands{ and }

\maketitle

\begin {abstract}
In this paper, we present a framework for resource allocations for multicast device-to-device (D2D) communications underlaying a cellular network. The objective is to maximize the sum throughput of active cellular users (CUs) and feasible D2D groups in a cell, while meeting a certain signal-to-interference-plus-noise ratio (SINR) constraint for both the CUs and D2D groups. We formulate the problem of power and channel allocation as a mixed integer nonlinear programming (MINLP) problem where one D2D group can reuse the channels of multiple CUs and the channel of each CU can be reused by multiple D2D groups. Distinct from existing approaches in the literature, our formulation and solution methods provide an effective and flexible means to utilize radio resources in cellular networks and share them with multicast groups without causing harmful interference to each other. A variant of the generalized bender decomposition (GBD) is applied to optimally solve the MINLP problem. A greedy algorithm and a low-complexity heuristic solution are then devised. 
The performance of all schemes is evaluated through extensive simulations.
Numerical results demonstrate that the proposed greedy algorithm can achieve close-to-optimal performance, and the heuristic algorithm provides good performance, though inferior than that of the greedy, with much lower complexity.
\end  {abstract}

\section{Introduction}

Device-to-Device (D2D) communication is a technology component for Long Term Evolution-Advanced (LTE-A)  of the Third Generation Partnership Project (3GPP) \cite{3gpp}. In D2D communication, cellular users (CUs) in close proximity can exchange information over a direct link rather than transmitting and receiving signals through a cellular base station (BS). D2D users communicate directly while remaining controlled under the BS. Compared to routing through a BS, CUs at close proximity can save energy and resources when communicating directly with each other.
Moreover, D2D users may experience high data rate and low transmission delay due to the short-range direct communication~\cite{X.lin}. Reducing the network load by offloading cellular traffic from the BS and other network components to a direct path between users is another benefit of D2D communication reduce the network load and increase its effective capacity. Other benefits and usage cases are discussed in~\cite{Shen}.

The majority of the literature in D2D communications uses the cellular spectrum for both D2D and cellular communications,also known as in-band  D2D~\cite{Doppler}. Generally, in-band D2D falls in two categories, underlay and overlay~\cite{Asadi}. Underlay in-band D2D can improve the spectrum efficiency of cellular networks by reusing cellular resources. Its main drawback lies in the interference caused by D2D users to cellular communications. Thus, efficient interference management and resource allocation are required to guarantee a target performance level of the cellular communication~\cite{Peng,Janis}.
In order to avoid this interference issue, it has also been proposed to dedicate part of the cellular resources to D2D communications in overlay in-band D2D. In this case, designing a resource allocation scheme is crucial to maximize the utilization of dedicated cellular resources~\cite{Pei}. Other works consider out-of-band instead of in-band D2D communications so that the cellular spectrum would not be affected by D2D communications~\cite{Asadi2}. 
Out-of-band D2D communication faces challenges in coordinating the communication over two different bands because usually D2D communication happens on a second radio interface (e.g., WiFi Direct and Bluetooth)~\cite{Fodor}. 

Most of the work in D2D resource allocation targets the unicast scenario where a single or multiple D2D pairs reuse the resources of CUs. In~\cite{Doppler}, the authors consider throughput maximization where by allowing D2D communication to underlay the cellular network, the overall throughput in the network can increase compared to a case where all D2D traffic is relayed by the cellular network. Some other work such as~\cite{Fodor},~\cite{Hyunkee2011} consider D2D communication reliability while guaranteeing a certain level of SINR or outage probability. 
The works in~\cite{Yu,M.zul,Feng} consider both throughput and reliability simultaneously. In~\cite{Yu}, throughput is maximized for a network with a single D2D pair and a single CU subject to spectral efficiency restrictions and energy constraints. There are few works for scenarios with multiple D2D users and CUs. For example, the quality-of-service (QoS) requirements for both CUs and D2D users have been investigated in~\cite{M.zul} and~\cite{Feng}. In~\cite{M.zul}, a heuristic algorithm has been proposed to solve the MINLP resource allocation problem that aims to decrease interference to the cellular network and maximize the total throughput. The authors in~\cite{Feng} present a framework of resource allocation for D2D communications underlaying cellular networks to maximize the overall network throughput of existing CUs and admissible D2D pairs while guaranteeing the QoS requirements for both CUs and D2D pairs. A scheme based on maximum weight bipartite matching  is proposed to determine a specific CU partner for each admissible D2D pair. 


Multicast D2D transmissions, where the same packets for a UE are sent to multiple receivers, are important for scenarios such as multimedia streaming, device discovery, and public safety. Specially, D2D multicast communications are required features in public safety services like police, fire and ambulance~\cite{3gpp}. 
Compared to communicating with each receiver separately in unicast D2D, multicast D2D
transmission reduces overhead and saves resources. However, unlike the more commonly studied unicast D2D (see e.g.~\cite{Yu}~\cite{Feng}), multicast D2D has its own challenges.
Within a multicast group, the data rates attainable at different receivers are different because of the diverse link conditions between each receiver and the transmitter. A common approach is to transmit at the lowest rate of all users within a group determined by the user with the worst channel condition. This assures that multicast services can be provided to all users. On the one hand, as all multicast users within a group receive the same data rate, the total sum rate grows with the number of active users of the group. 
On the other hand, the lowest transmission rate typically decreases as the number of users increases since it is based on the user with the Least Channel Gain (LCG)~\cite{Afolabi}.

As discussed in~\cite{Afolabi} there are lots of works in multicast scheduling and resource allocation for OFDMA-based systems. They can be broadly classified into two types: single-rate and multi-rate transmissions. In single-rate broadcast, the BS transmits to all users in each multicast group at the same rate irrespective of their non-uniform achievable capacities, whereas in multirate broadcast, the BS transmits to each user in each multicast group at different rates based on what each user can handle. All of the works mentioned in~\cite{Afolabi} targeted cellular networks where the multicast transmitter is the BS. However, in multicast D2D, UEs are multicast transmitters and the QoS requirements for both the D2D links and the cellular links should be satisfied. 

The problem of resource management for D2D multicast communication was first addressed in our previous work~\cite{Meshgi15}. In~\cite{Meshgi15} we formulate the power and channel allocation problem for D2D multicast communication for a special case where each D2D group can reuse the channel of one CU and the channel of each CU can be reused by at most one D2D group. The optimal solution is found using maximum weight
bipartite matching algorithm and a low-complexity heuristic algorithm
is also proposed. Moreover, we adapt the heuristic scheme in~\cite{M.zul} for multicast D2D and compare it against our scheme and show that our proposed heuristic has a superior performance.

In this paper, we consider multicast D2D communications underlaying cellular networks and present a joint power and channel allocation scheme to maximize the total throughput of all CUs and D2D groups within a cell. We formulate the general problem of power and channel allocation as an MINLP where one D2D group can reuse the channels of multiple CUs and the channel of each CU  can be reused by multiple D2D groups. To guarantee the QoS requirements for both CUs and D2D groups, a minimum SINR constraint is imposed.
A variant of the generalized bender decomposition (GBD) is applied to optimally solve the MINLP problem. We further propose an exact solution to a special case of the general problem. Specifically, inspired by the work in~\cite{Feng}, we use the maximum weight bipartite matching algorithm for the case where each D2D group can reuse the channel of at most one CU and each CU can share their resources with at most one D2D group. Next, we propose a greedy algorithm with a somewhat high complexity but very close-to-optimal performance.
A low-complexity heuristic solution is then devised which trades computation complexity with performance. This heuristic algorithm is an extension to the heuristic algorithm presented~\cite{Meshgi15} for the general scenario. 

The remainder of the paper is organized as follows. In Section~\ref{sec:model}, the system model is described and the problem of power and channel allocation for underlay multicast D2D communication is formulated. Section~\ref{sec:GBD} describes the generalized bender decomposition method to solve the general problem. The matching-based optimal resource allocation for one special case is presented in Section~\ref{sec:matching}, and the greedy and the heuristic algorithms are presented in Section~\ref{sec:algorithms}. Numerical results are demonstrated in Section~\ref{sec:results}, and Section~\ref{sec:conclusions} concludes the paper.


\section{System Model and Problem Formulation}
\label{sec:model}

We study resource allocation for multicast D2D communcations underlaying uplink (UL) transmissions in LTE networks. UL resource sharing is considered since reusing downlink resources is more difficult and less effective than reusing uplink resources in the worst case of a fully loaded cellular network, as demonstrated in~\cite{Doppler2}.
Consider $K$ groups of multicast D2D users coexisting with $M$ CUs. 
We assume a fully loaded cellular network scenario. That is, there are $M$ channels, each occupied by one CU. We use $ m \in \mathcal{M}=\{1,2,\ldots,M\}$ to index both the $m$th CU and the channel it occupies, and $k \in \mathcal{K}=\{1,2,\ldots,K\}$ to index the $k$th D2D group. We consider a single cell scenario and assume that advanced intercell interference mitigation is applied on top of our scheme.
Within a D2D group, there is only one user that multicasts messages to the remaining users. Each D2D user only belongs to one D2D group. We use $\mathcal{D}_k$ to represent the set of D2D receivers in the $k$th multicast group, and $|\mathcal{D}_k|$ is the total  number
of receivers in the group. As a special case, when $|\mathcal{D}_k|=1$, the scenario becomes unicast. 

Define a set of binary variables $y_{km}$ with $y_{km}=1$ if the $k$th D2D group reuses channel $m$, and $y_{km}=0$ otherwise. 
In the general case, each D2D group splits its multicast traffic among maximally $C_1$ channels where $C_1 \le M$, and each channel can be reused by at most $C_2$ D2D groups where $C_2 \le K$. That is,
\begin{align}
 \label{eq:Ykm1}
 && \sum_{m=1}^M y_{k,m} \le C_1, \ \forall k \in \mathcal{K} \\
 \label{eq:Ykm2}
 && \sum_{k=1}^K y_{k,m} \le C_2, \ \forall m \in \mathcal{M}.
\end{align}

The channel quality of receiver $d$ in the $k$th D2D group at channel $m$ is given by

\begin{equation}
\label {eq:3}
\beta^{D2D}_{k,m,d} = \frac {{G^{D2D}_{k,m,d}}} {P_{noise}+P_m^{Cell} G^{C2D}_{k,m,d}+ \sum_{k'\ne k} P^{D2D}_{k',m} G^{D2D}_{k,k',d}},
\end{equation}
where $P_{noise}$ is the aggregate power of background noise, $G^{D2D}_{k,m,d}$ is the link gain to D2D receiver $d$ from the D2D transmitter in group $k$ at channel $m$, $G^{C2D}_{k,m,d}$ is the link gain from CU $m$ to D2D receiver $d$ in group $k$, $P_m^{Cell} $ is the transmission power of CU $m$, $P^{D2D}_{k,m}$ is the transmission power of the $k$th D2D group transmitter at channel $m$, and $G^{D2D}_{k,k',d}$  the link gain from the transmitter at D2D group $k'$ to receiver $d$ at D2D group $k$.

For the $k$th D2D group, its transmission condition in channel $m$ is determined by the receiver with the worst condition. Define
\begin{align}
\label{eq:betad2d}
 \beta^{D2D}_{k,m} = \min_{d\in\mathcal{D}_k} \beta^{D2D}_{k,m,d}.
\end{align}
Then, the normalized transmission rate (bit/s/Hz) of the $k$th D2D group is given by
\begin{equation}
\label {eq:D2Drate}
 r^{D2D}_{k}= \sum_{m=1}^M y_{k,m}\log_2(1+P^{D2D}_{k,m} \beta^{D2D}_{k,m}).
\end{equation}

The aggregate transmission rate of the $k$th D2D group is given by
\begin{equation}
\label {eq:3}
R^{D2D}_{k}= |\mathcal{D}_k| r^{D2D}_{k}.
\end{equation}
For CU $m$, its channel quality is given by 
\begin{equation}
\label  {eq:betacell}
\beta^{Cell}_{m}=\frac{G^{Cell}_{m}} {P_{noise}+ \sum_{k=1}^K y_{k,m} P^{D2D}_{k,m} G^{D2C}_{k,m}},
\end{equation}
where $G^{Cell}_{m}$ is the link gain of CU $m$ to the cellular base station, and $G^{D2C}_{k,m}$ is the link gain from the $k$th D2D transmitter to the cellular base station at channel $m$. Therefore,
the normalized transmission rate for CU $m$ is 
\begin{equation}
\label{eq:CUrate}
R^{Cell}_{m}=   \log_2(1+P^{Cell}_m \beta^{Cell}_{m}).
\end{equation}

A threshold is set for the SINR of each D2D group and CU transmission. For the $k$th D2D group, 
\begin{align}
\label{eq:SNRD}
 P^{D2D}_{k,m} \beta^{D2D}_{k,m} \ge y_{k,m}\gamma^{D2D}_{th},
\end{align}
and for CU $m$,
\begin{align}
\label{eq:SNRC}
P^{Cell}_m \beta^{Cell}_{m} \ge \gamma^{Cell}_{th}.
\end{align}

Given these SINR threshold constraints, we can approximate the capacity in higher SINR cases by removing the term ``1'' from the logarithm functions in both~\eqref{eq:D2Drate} and~\eqref{eq:CUrate}.
The maximum power constraints for CUs and D2D groups, respectively, are given by
\begin{equation}
\label  {eq:Pc}
P^{Cell}_m \le P^{Cell}_{\max} , \ \forall  m \in \mathcal{M},\\
\end{equation}
and
\begin{equation}
\label  {eq:Pd}
 \sum_{m=1}^M P^{D2D}_{k,m} \le P^{D2D}_{\max} , \ \forall k \in \mathcal{K}.
\end{equation}

The objective is to maximize the aggregate data transmission rate of all the D2D groups and CUs. 
Combining~\eqref{eq:Ykm1} -- \eqref{eq:Pd}, we formulate the joint power control and channel allocation problem to maximize the sum throughput of multicast D2D groups and cellular users as follows,
  {
 \allowdisplaybreaks
\begin{eqnarray}
\label {eq:40}
  &\mbox{ P1.} & \max  \left( \sum_{k=1}^K R^{D2D}_k + \sum_{m=1}^M R^{Cell}_m  \right)\\
 & \mbox{s.t.} &  R^{D2D}_{k} =   \sum_{m=1}^M y_{k,m} |\mathcal{D}_k|    \log_2(P^{D2D}_{k,m} \beta^{D2D}_{k,m}), \forall k \in \mathcal{K}, m \in \mathcal{M},\\ 
  &&R^{Cell}_m = \sum_{k=1}^K \log_2\left(P^{Cell}_m \beta^{Cell}_{m}\right),  \ \forall m \in \mathcal{M}, \\
&&   \beta^{D2D}_{k,m} \le \beta^{D2D}_{k,m,d}, \ \forall k \in \mathcal{K}, m \in \mathcal{M}, \ d\in\mathcal{D}_k,\\
\label  {Y3}
&& y_{k,m} \in \{0,1\},  \ \forall k \in \mathcal{K}, m \in \mathcal{M},\\ 
 &&  \mbox{Constraints}~\eqref{eq:Ykm1},~\eqref{eq:Ykm2},~\eqref{eq:betacell},~\eqref{eq:SNRD},~\eqref{eq:SNRC},~\eqref{eq:Pc},~\eqref{eq:Pd} \nonumber.
\end{eqnarray}

Table~\ref{param} lists all the parameters and variables used in the problem formulation.

Clearly, P1 is a Mixed Integer Nonlinear Programming (MINLP) problem.
In general, MINLP problems are NP-hard and thus no efficient polynomial-time solutions exist. In the general case, when $C_1$ and $C_2$ are arbitrary values, we will  use GBD~\cite{GBD} to solve the problem in the next section.

Based on the values of $C_1$ and $C_2$, several special cases exist. For example,  when $C_1=1$ and $C_2=1$, each D2D group can reuse the channels of at most one CU and each CU can share their channels with at most one D2D group. Another special case of interest is when $C_1=1$. In this case, to increase the spectrum utilization, we allow each D2D group to reuse the resources of multiple CUs, but each CU cannot share its resource with more than one D2D group. Here, there is no interference between D2D groups and this setting is useful when the number of D2D groups is much less than the number of CUs. All the special cases can be resolved via GBD. However, it turns out that polynomial algorithm can be devised when $C_1=1$ and $C_2=1$ as will be discussed in Section~\ref{sec:matching}.

\begin{table}[h]
\centering
\caption{Table of notations}
\begin{tabular}{ | l | p{13.5cm} |}
\hline
\rowcolor[HTML]{9B9B9B} 
\textbf{Notation} & \textbf{Description}\\ \hline
$\mathcal{M}$                        & Set of cellular users (CU)             \\ \hline
$\mathcal{K} $          		    & Set of D2D  groups                        \\ \hline
$\mathcal{D}_k$          		    & Set of receivers  in $k$th D2D group \\ \hline
$\mathcal{A} $          		    & Set of admissible or successful D2D  groups                        \\ \hline
$y_{k,m}$                              & Binary variable, =1 if $k$th D2D group reuses CU $m$'s channel, and =0 otherwise   \\ \hline
$C_1$                        	& Maximum number of channels to be reused by each D2D group    \\ \hline
$C_2$                        	& Maximum number of D2D groups sharing each CU channel    \\ \hline
$P_{noise}$                           & Aggregate power of background noise                 \\ \hline
$G^{D2D}_{k,m,d}$        &Link gain to D2D receiver $d$ from the D2D transmitter in group $k$ at channel $m$   \\ \hline
$G^{C2D}_{k,m,d}$ 	      &Link gain from CU $m$ to D2D receiver $d$ in group $k$   \\ \hline
$G^{D2D}_{k,k',d}$       &Link gain from the transmitter at D2D group $k'$ to receiver $d$ at D2D group $k$   \\ \hline
$G^{Cell}_{m}$       &Link gain of CU $m$ to the cellular base station   \\ \hline
$G^{D2C}_{k,m}$       &Link gain from the $k$th D2D transmitter to the cellular base station at channel $m$  \\ \hline
$P^{D2D}_{k,m}$         &Transmission power of the $k$th D2D group transmitter at channel $m$                  \\ \hline
$P_m^{Cell} $                         &Transmission power of CU $m$                          \\ \hline
$\beta^{D2D}_{k,m,d} $           &Channel quality of receiver $d$ in the $k$th D2D group at channel $m$               \\ \hline
$\beta^{Cell}_{m}$                      & Channel quality of CU $m$                             \\ \hline
$R^{D2D}_{k}$                      &Normalized transmission rate of the $k$th D2D group       \\ \hline
$R^{Cell}_{m}$                      &Normalized transmission rate for CU $m$       \\ \hline
$R^{sum}$                      &The summation of D2D and cellular throughput       \\ \hline
$\gamma^{D2D}_{th}$                      &SINR threshold for all D2D groups       \\ \hline
$\gamma^{Cell}_{th}$                      &SINR threshold for all CUs     \\ \hline
$f_i(| \mathcal{D_K}|)$                     &The complexity of solving problem Pi    \\ \hline
\end{tabular}
\label{param}
\end{table}

\section{Generalized Bender Decomposition}
\label{sec:GBD}

The MINLP problem in P1 has the special property that when the binary variables ($y_{k,m}$'s) are fixed, the problem becomes a geometric programming problem with continuous variables ($P^{D2D}_{k,m}$'s and $P^{Cell}_m$'s), which can be transformed to a convex problem. This allows us to use GBD~\cite{GBD} to solve the problem efficiently with proper transformation.

Let $\mathbf{X} =[P^{D2D}_{k,m},P^{Cell}_{m}, R^{D2D}_{k}, R^{Cell}_{m},\beta^{D2D}_{k,m}, \beta^{Cell}_{m},k \in \mathcal{K},m \in \mathcal{M}] $ represent the set of all continuous variables and $\mathbf{Y}=[y_{k,m},k \in \mathcal{K},m \in \mathcal{M}]$ represent the binary variables.
We modify the constraints in problem P1 to separate binary variables $y_{km}$ from the continuous variables in $\mathbf{X}$ and make the problem linear in terms of $y_{k,m}$'s when the continuous variables are fixed. Problem P1 can be transformed as
  {
 \allowdisplaybreaks
\begin{eqnarray}
\label {eq:40}
  &\mbox{ P2.} & f(\mathbf{X}, \mathbf{Y} ) =\max  \left( \sum_{k=1}^K R^{D2D}_k + \sum_{m=1}^M R^{Cell}_m  \right)\\
 & \mbox{s.t.} & R^{D2D}_{k} \le  \sum_{m=1}^M  |\mathcal{D}_k|    \log_2(P^{D2D}_{k,m} \beta^{D2D}_{k,m}) +C(1-y_{k,m}), \forall k \in \mathcal{K}, m \in \mathcal{M},\\ 
  &&   R^{D2D}_{k} \le   Cy_{k,m}, \forall k \in \mathcal{K}, m \in \mathcal{M},\\
  &&R^{Cell}_m \le \sum_{k=1}^K \log_2\left(P^{Cell}_m \beta^{Cell}_{m}\right),  \ \forall m \in \mathcal{M}, \\
  \label  {C2}
   &&  \beta^{Cell}_{m} \le \frac{G^{Cell}_{m}} {P_{noise}+  \sum_{k=1}^KP^{D2D}_{k,m} G^{D2C}_{k,m}},\ \forall m \in \mathcal{M},\\
   \label  {B2}
&& \beta^{D2D}_{k,m} \le \frac {{G^{D2D}_{k,m,d}}} {P_{noise}+P_m^{Cell} G^{C2D}_{k,m,d}+ \sum_{k'\ne k}P^{D2D}_{k',m} G^{D2D}_{k,k',d}},\ \forall k \in \mathcal{K}, m \in \mathcal{M}, \ d\in\mathcal{D}_k, \\
  && \textstyle \frac{P^{D2D}_{k,m}}{P^{D2D}_{\max} } \le y_{k,m} +\epsilon \le CP^{D2D}_{k,m} , \ \forall k \in \mathcal{K} , m \in \mathcal{M},\\ 
 &&  \mbox{Constraints}~\eqref{eq:Ykm1},~\eqref{eq:Ykm2},~\eqref{eq:betacell},~\eqref{eq:SNRD},~\eqref{eq:SNRC},~\eqref{eq:Pc},~\eqref{eq:Pd}, ~\eqref{Y3} \nonumber.
  \label {eq:50}
\end{eqnarray}
%
where $C$ is a very large number and $\epsilon>0 $ is a very small number. 

The basic idea of GBD is to decompose the original MINLP problem into a primal problem and a master problem, and solve them iteratively. The primal problem corresponds to the original problem with fixed binary variables. Solving this problem provides the information about the lower bound and the Lagrange multipliers corresponding to the constraints. The master problem is derived through nonlinear duality theory using the Lagrange multipliers obtained from the primal problem. The solution to the master problem gives the information about the upper bound as well as the binary variables that can be used in the primal problem in next iteration. When the upper bound meets the lower bound, the iterative process converges.

\subsection {Primal problem}

The primal problem results from fixing the ${y_{k,m}}$ variables to a particular 0-1 combination denoted by $y^{(i)}_{k,m}$,  where $i$ stands for the iteration counter. The formulation for the primal problem at iteration $i$ is
given by
  {
 \allowdisplaybreaks
\begin{eqnarray}
\label {eq:P3}
  &\mbox{P3.} & f(\mathbf{X}, \mathbf{Y^{(i)}} ) = \max  \left(\sum_{k=1}^K R^{D2D}_k + \sum_{m=1}^M R^{Cell}_m  \right) \\
 & \mbox{s.t.} & R^{D2D}_{k} \le  \sum_{m=1}^M  |\mathcal{D}_k|    \log_2(P^{D2D}_{k,m} \beta^{D2D}_{k,m}) +C(1-y^{(i)}_{k,m}), \forall k \in \mathcal{K}, m \in \mathcal{M},\\ 
  &&   R^{D2D}_{k} \le   Cy^{(i)}_{k,m}, \forall k \in \mathcal{K}, m \in \mathcal{M},\\
  &&R^{Cell}_m \le \sum_{k=1}^K \log_2\left(P^{Cell}_m \beta^{Cell}_{m}\right),  \ \forall m \in \mathcal{M}, \\
   && \beta^{Cell}_{m} \le \frac{G^{Cell}_{m}} {P_{noise}+  \sum_{k=1}^KP^{D2D}_{k,m} G^{D2C}_{k,m}},\ \forall m \in \mathcal{M},\\
&&\beta^{D2D}_{k,m} \le \frac {{G^{D2D}_{k,m,d}}} {P_{noise}+P_m^{Cell} G^{C2D}_{k,m,d}+ \sum_{k'\ne k}P^{D2D}_{k',m} G^{D2D}_{k,k',d}}, \ \forall k \in \mathcal{K}, m \in \mathcal{M}, \ d\in\mathcal{D}_k, \\
  && \textstyle \frac{P^{D2D}_{k,m}}{P^{D2D}_{\max} } \le y^{(i)}_{k,m} +\epsilon \le CP^{D2D}_{k,m} , \ \forall k \in \mathcal{K} , m \in \mathcal{M},\\ 
  &&P^{D2D}_k \beta^{D2D}_{k,m} \ge y^{(i)}_{k,m}\gamma^{D2D}_{th}, \ \forall k \in \mathcal{K}, m \in \mathcal{M},\\ 
 &&  \mbox{Constraints}~\eqref{eq:SNRC},~\eqref{eq:Pc},~\eqref{eq:Pd},~\eqref{C2},~\eqref{B2}\nonumber.
\end{eqnarray}}

Since the optimal solution to this problem is also a feasible solution to problem P1, the optimal value $ f(\mathbf{X^*}, \mathbf{Y^{(i)}} ) $ provides a lower bound to the original problem.
In general, not all choices of binary variables lead to a feasible primal problem. Therefore, for a given choice of ${y_{k,m}}$'s, there are two cases for primal problem P3: feasible problem and infeasible problem. In the following, we consider each of these cases.
\begin{itemize}
 \item Feasible Primal: If the primal problem at iteration $i$ is feasible, then its solution provides information on the transmission power of D2D and cellular transmitters, $ f(\mathbf{X^*}, \mathbf{Y^{(i)}})$, and the optimal multiplier vectors, $\lambda^k_q$, $q=1,2,\ldots,Q$ for the $Q$ inequality constraints in Problem P3. Subsequently, using this information we can formulate the Lagrange function for all inequality constraints  $G_q(\mathbf{X}, \mathbf{Y^{(i)}} ) \le 0 $ for $q=1,2,\ldots,Q$ as
  {
 \allowdisplaybreaks
\begin{eqnarray}
\label {eq:P6}
  &&L(\mathbf{X}, \mathbf{Y^{(i)}} ,\lambda^{(i)})= f(\mathbf{X}, \mathbf{Y^{(i)}}) + \sum_{q=1}^Q  {\lambda^{(i)}_q G_q(\mathbf{X}, \mathbf{Y^{(i)}} )},
\end{eqnarray}}
%

where $\lambda^{(i)}=[\lambda^{(i)}_q, q=1,2,\ldots,Q]$.

\item Infeasible Primal:  If the primal problem is infeasible, to identify a feasible point we can formulate an $l_1$-minimization problem as
  {
 \allowdisplaybreaks
\begin{eqnarray}
\label {eq:P6}
  &\mbox{P3.1.} & \min \sum_{q=1}^Q \alpha_q\\
 & \mbox{s.t.} &  G_q(\mathbf{X}, \mathbf{Y^{(i)}} )\le \alpha_q , q=1,2,...,Q,\\ 
 &&   \alpha_q\ge 0 , q=1,2,...,Q.
\end{eqnarray}}
Note that if $\sum_{q=1}^Q \alpha_q=0$, then P3 is feasible. Otherwise,
the solution to this feasibility problem (FP) provides information on the Lagrange multipliers, which are denoted as $\bar\lambda^{(i)}_q$; the Lagrange function resulting from the feasibility problem at iteration $i$ can be defined as
  {
 \allowdisplaybreaks
\begin{eqnarray}
\label {eq:P6}
  && \bar L(\mathbf{X}, \mathbf{Y^{(i)}} ,\bar\lambda^{(i)}) = \sum_{q=1}^Q \bar\lambda_q^{(i)}( G_q (\mathbf{X}, \mathbf{Y^{(i)}} ) -\alpha_q).
\end{eqnarray}}

\end{itemize}

\subsection{Master Problem}
The master problem is derived from the non-linear duality theory~\cite{GBD}.
{
 \allowdisplaybreaks
\begin{eqnarray}
  &\mbox{P4.} &\max_{\mathbf{Y}^{(i)} } \eta\\
  & \mbox{s.t.} & \eta \le   \sup _{\mathbf{X}}L(\mathbf{X}, \mathbf{Y^{(i)}},\lambda^{(i)}),  \ \forall \lambda^{(i)} \ge 0,  \\
 && \inf _{\mathbf{X}} \bar L(\mathbf{X}, \mathbf{Y^{(i)}},\bar\lambda^{(i)}) \le 0,  \ \forall \bar\lambda^{(i)} \in \Lambda, \\
&&  \mbox{Constraints}~\eqref{eq:Ykm1},~\eqref{eq:Ykm2},~\eqref{Y3},
\end{eqnarray}}
where 
\begin{equation}
\label  {eq:Lambda}
\Lambda= \{ \bar\lambda_q \ge 0,  \sum_{q=1}^Q \bar\lambda_q=1\}.
\end{equation}

The master problem P4  is similar to the original problem P2, but has two inner optimization
problems which need to be considered for all $\lambda$ and $\overline\lambda$
 obtained from the primal problem in every iteration. Therefore, it has a very large number of constraints. 
Because of the separability of  binary variables $\mathbf{Y}$ and continuous variables $\mathbf{X}$, and the linearity with regard to $\mathbf{Y}$, we can adopt Variant 2 of GBD (V2-GBD) in ~\cite{GBD}. It is proven in~\cite{GBD} that under the conditions for V2-GBD, the Lagrange function evaluated at the solution of the corresponding primal is a valid under-estimator of the inner optimization problem in P4. Therefore, the relaxed master problem can be formulated as,
 
 {
 \allowdisplaybreaks
\begin{eqnarray}
  &\mbox{P5.} &\max_{\mathbf{Y}^{(i)} } \eta\\
  & \mbox{s.t.} & \eta \le  L(\mathbf{X}, \mathbf{Y^{(i)}},\lambda^{(i)}),  \ \forall \lambda^{(i)} \ge 0,  \\
 && \bar L(\mathbf{X}, \mathbf{Y^{(i)}},\bar\lambda^{(i)}) \le 0,  \ \forall \bar\lambda^{(i)} \in \Lambda,  \\
&&  \mbox{Constraints}~\eqref{eq:Ykm1},~\eqref{eq:Ykm2},~\eqref{Y3}.
\end{eqnarray}}

The relaxed problem provides an upper bound to the master problem and can be used to generate the primal problem in the next iteration.The same procedure is then repeated until convergence.
Over the iterations, the sequence of upper bounds are nonincreasing and the set of lower bounds are nondecreasing. The two sequences are proven to converge and the algorithm will stop at the optimal solution within a finite number of iterations~\cite{Zheng}. Algorithm~\ref{alg:GBD} summarizes the GBD procedure.

\begin{algorithm}{}
\caption{GBD Algorithm}
\label{alg:GBD}
\begin{algorithmic}[1]
\STATE {First iteration, $i=1$}
\STATE {Select an initial value for $\mathbf{Y}^{(i)}$, which makes the primal problem feasible.}
\STATE {Solve the primal problem in P3 and obtain the Lagrange function  }
\STATE {$UBD^{(i)}=\infty$, $LBD^{(i)}=0$}
\WHILE {$ UBD^{(i)} - LBD^{(i)} > 0 $}
	\STATE{$i=i+1$}
	\STATE { Solve the relaxed master problem P5 to optain $\eta^*$ and $\mathbf{Y}^{*}$}
	\STATE {Set $UBD^{(i)}=\eta^*$}
	\STATE {Solve the primal problem P3 with fixed $\mathbf{Y}^{(i)}=\mathbf{Y}^{*}$}
		\IF {the primal problem is feasible}
			\STATE{Obtain optimal solution $\mathbf{X^*} $ and the Lagrange function $L(\mathbf{X}, \mathbf{Y^{(i)}},\lambda^{(i)})$}
			\STATE{Set $LBD^{(i)}= \max (LBD^{(i-1)}, f^{(i)}(\mathbf{X^*}, \mathbf{Y^{(i)}} ))$}
		\ELSE  
			\STATE{Solve the feasibility-check problem P3.1 to obtain the optimal solution $\mathbf{X^*} $ and the Lagrange function $ \bar L(\mathbf{X}, \mathbf{Y^{(i)}},\bar\lambda^{(i)})$}
		\ENDIF
\ENDWHILE
\end{algorithmic}
\end{algorithm}


\section {Matching-based Optimal Resource Allocation}
\label{sec:matching}

In this section, we consider the MINLP problem in P1 for the special case  $C_1=1$ and $C_2=1$. This case can be cast as a bipartite matching problem and thus can be solved polynomially. To formulate the bipartite problem, we divide P1 into two subproblems. 
In the first step, for each D2D group $k$ and each CU $m$, we find their transmission power so that the sum throughput of the D2D group and the CU is maximized. If this problem is feasible, D2D group $k$ is allowed to reuse the channel of CU $m$ and is marked as a candidate partner in the second step; otherwise group k is excluded  from the list of feasible partners.
The second step is then to find the best CU partner for each D2D group among all feasible candidates so that the total throughput of all D2D groups and CUs is maximized. 


\subsubsection{Feasibility check and power allocation}
In order to determine whether D2D group $k$ can reuse channel $m$ and to find the transmission power of the feasible D2D group and CU,we have problem P6 as follows:
 {
 \allowdisplaybreaks
\begin{eqnarray}
\label {eq:P2}
 &\mbox{P6.} & \max \left( R^{D2D}_{k,m} + R^{Cell}_{k,m} \right)\\
 &\mbox{s.t.} &  R^{D2D}_{k,m} =    |\mathcal{D}_k|    \log_2(P^{D2D}_{k,m} \beta^{D2D}_{k,m}), \\
 && R^{Cell}_{k,m} =\log_2\left( P^{Cell}_{m} \beta^{Cell}_{m} \right), \\
 && P^{D2D}_{k,m} \beta^{D2D}_{k,m} \ge \gamma^{D2D}_{th},  \\
 && P^{Cell}_{m} \beta^{Cell}_{m} \ge  \gamma^{Cell}_{th},  \\
   &&  \beta^{Cell}_{m}=  \frac{G^{Cell}_{m}} {P_{noise}+ P^{D2D}_{k,m} G^{D2C}_{k,m}},\\
&& \beta^{D2D}_{k,m} \le  \frac {G^{D2D}_{k,m,d}} {P_{noise}+  P^{Cell}_{m} G^{C2D}_{k,m,d}}, \ \forall d\in\mathcal{D}_k \\
&& P^{Cell}_{m} \le P^{Cell}_{\max} ,  \\
&& \sum_{m=1}^M P^{D2D}_{k,m} \le P^{D2D}_{\max}  .
\end{eqnarray}}

P6 is a reduced version of P1 by limiting it to only one D2D group and one CU with the objective of maximizing their sum throughput. Clearly, P6 is a geometric programming problem and can be transformed to a convex optimization problem using geometric programming techniques~\cite{GP}.
We solve problem P6 for all $k$ and $m$ pairs. Define a candidate channel set $\mathcal{C}_k$ for D2D group $k$. If the problem is feasible, D2D group $k$ is admissible to channel $m$ (i.e., eligible to use channel $m$), then $m$ is added to $\mathcal{C}_k$. 
For $m\in\mathcal{C}_k$, denote the optimal throughput for the $k$th D2D transmitter and the $m$th CU as $R^{*D2D}_{k,m}$ and $ R^{*Cell}_{k,m}$, respectively, and the optimal sum throughput as $R^{sum}_{k,m}=R^{*D2D}_{k,m}+R^{*Cell}_{k,m}$. For $m\notin \mathcal{C}_k$, we set $R^{*D2D}_{k,m}=0$, $ R^{*Cell}_{k,m}=\log_2\left(\frac{P^{Cell}_{\max} G^{Cell}_{m}} {P_{noise}} \right)$, and thus $R^{sum}_{k,m}= R^{*Cell}_{k,m}$.


\subsubsection{Maximizing total throughput}

Given the maximum achievable throughput for each D2D group when reusing each cellular channel, to find the optimal channel allocation that maximizes the total throughput we have,

       {
 \allowdisplaybreaks
\begin{eqnarray}
  &\mbox{P7.} &\max_{y_{k,m}} \sum_{k=1}^K \sum_{m=1}^M y_{k,m} R^{sum}_{k,m} \\
  & \mbox{s.t.} &  \sum_{k=1}^K y_{k,m} \le 1 ,\ \forall  m \in \mathcal{M} \\
&&
   \sum_{m=1}^M y_{k,m} \le 1 , \ \forall k \in \mathcal{K} \\
&&
y_{k,m} \in \{0,1\},  \ \forall k \in \mathcal{K}, \  m \in \mathcal{M}.
\end{eqnarray}}
P7 is in effect the maximum weight bipartite matching problem,  where the D2D groups and the cellular channels are two groups of vertices in the bipartite graph, and the edge connecting D2D group $k$ and  channel $m$ has a weight $R^{sum}_{k,m}$.
The Hungarian algorithm~\cite{Khun} can be used to solve the bipartite matching problem in polynomial time.
%

To determine the computational complexity, consider $M \ge K$ and the complexity of solving P6 is a function of the size of each D2D group, i.e. $f_6(| \mathcal{D_K}|)$. Therefore, the time complexity of the matching-based optimal resource allocation is $\textbf {O}(M\times K \times f_6(| \mathcal{D_K}|))+ \textbf {O}(M^3)$ , where the first and second terms correspond to the computation time in the first and second steps, respectively.


\section{ Greedy and Heuristic Channel allocation algorithms}
\label{sec:algorithms}

The MINLP problem in P1 is an NP-hard problem and the computation complexity is exponential in the worst case. In other words, GBD may converge in an exponential number of iterations. In this section we first propose a greedy algorithm and then a  heuristic solution to the general MINLP problem in P1.

\begin{algorithm}{}
\caption{Greedy algorithm}
\label{alg:greedy}
\begin{algorithmic}[1]
\STATE {$\mathcal{M}$: Set of cellular users}
\STATE {$\mathcal{K}$: Set of all D2D groups}
\STATE{$ e_{k,m}=1, \ \forall k \in \mathcal{K},  m \in \mathcal{M} $}
\STATE{$Y=[y_{k,m}|\mbox{ } y_{k,m}=0, \ \forall k \in \mathcal{K},  m \in \mathcal{M} $]}
\STATE {$\mathcal{S}=\emptyset$}
\WHILE {$  \sum_{k=1}^K  \sum_{m=1}^M e_{k,m} \ge 1 $ }
\STATE{$E=[e_{k,m}|\mbox{ } e_{k,m}=1, \ \forall k \in \mathcal{K},  m \in \mathcal{M} $]}
\STATE{$ T^{sum}_{k,m}=\sum_{m'=1}^M \log_2\left(\frac{P^{Cell}		_{\max} G^{Cell}_{m'}} {P_{noise}} \right), \ \forall k \in \mathcal{K}, m \in \mathcal{M} $}
		\FOR  {$ \textbf{each} \mbox{ }e_{k,m} \in E$}
			\STATE{$y_{k,m}=1$}
			\IF { $(k,m)$ is \textit{Admissible} }
				\STATE{Solve P3 to find $P^{D2D}_{k',m'} $ and $P^{Cell}_{m'},  \ \forall (k',m') \in \mbox{ }[\mathcal{S} \mbox{ } \cup \mbox{ }  (k,m)]$ }
				\IF { P3 is feasible }
					\STATE{$T^{sum}_{k,m}= \sum_{(k',m')\in [\mathcal{S} \mbox{ } \cup 	\mbox{ }  (k,m)]} y_{k',m'} |	\mathcal{D}_{k'}|\log_2(P^{D2D}_{k',m'} \beta^{D2D}_{k',m'})+ 							 \sum_{m'=1}^M \log_2\left(P^{Cell}_{m'} \beta^{Cell}_{m'}\right)$}
				\ELSE
					\STATE {$e_{k,m}=0$}
  				\ENDIF
  			\ELSE
  				\STATE {$e_{k,m}=0$}
  			\ENDIF
			\STATE{$y_{k,m}=0$}
		\ENDFOR
		\STATE{$(k^*, m^*)=\arg\max_{\forall (k,m) }T^{sum}_{k,m}$}
		\STATE{$y_{k^*,m^*}=1$}
		\STATE {$e_{k^*,m^*}=0$}
		\STATE { $\mathcal{S} = \mathcal{S} \mbox{ } \cup \mbox{ }  (k^*,m^*)$}
\ENDWHILE

\end{algorithmic}
\end{algorithm}

\subsection{Greedy algorithm}

Algorithm~{\ref{alg:greedy}} shows the greedy resource allocation algorithm. The key idea of the greedy algorithm is that, in each iteration, it selects a CU and D2D group pair that maximizes the resulting sum throughput of all selected pairs. The algorithm terminates when there is no more pair that can be included. 

In this algorithm, we first initialize all edges of a $K \times M$ bipartite graph ,$e_{k,m}$, to one in line 3.  The $K \times M$ assignment matrix $\bold{Y}$ is initialized to zero. $\mathcal{S}$ is the set of  selected CU and D2D pairs that maximize the sum throughput and initialize to zero at first. Matrix $E$ includes all edges ($e_{k,m}$) with the value of one.  The inner loop (lines 8-23) finds the sum throughput, $T^{sum}_{k,m}$, of all pairs in set $\mathcal{S}$ after an admissible pair $(k,m)$ is added to $\mathcal{S}$. In line 10, to find if $(k,m)$ is \textit{admissible}, the algorithm checks constraints~\eqref{eq:Ykm1} and~\eqref{eq:Ykm2} for a given $(k,m)$ pair. If either of these constraints is violated for the current $(k,m)$, the procedure sets $e_{k,m}$ and $y_{k,m}$ to zero and moves to the next pair. Otherwise, the algorithm solves problem P3 and finds $T^{sum}_{k,m}$.
In the outer loop, the pair $(k^*, m^*)$ that maximizes $T^{sum}_{k,m} \ \forall (k,m) \in \mathcal{S}$ (line 24) is found and removed from $E$. 
The outer loop is iterated until $e_{k,m}=0, \ \forall k \in \mathcal{K}$ and $m \in \mathcal{M} $.

Since a total of $\min\{M\times C_2, K\times C_1\}$ pairs can be found in the procedure, and in each iteration of the outer loop, only one such pair can be added, the computational complexity of the greedy algorithm is $\textbf {O}(\min\{M\times C_2, K\times C_1\}\times K \times M\times f_3(|\mathcal{D_K}|))$, where $f_3(| \mathcal{D_K}|)$ is the complexity of solving P3 as a function of the size of each D2D group. The high complexity of the greedy algorithm mainly arises from the need to solve the optimization problem up to $K\times M$ times to find the best pair in each iteration.

\begin{algorithm}{}
\caption{Heuristic algorithm}
\label{alg:hur}
\begin{algorithmic}[1]
\STATE {$\mathcal{M}$: List of cellular users in decreasing order of $G^{Cell}_{m}$}
\STATE {$\mathcal{K}$: List of all D2D groups}
\STATE {$G^{C2D}_{m,k}= \min_{d\in\mathcal{D}_k} G_{k,m,d}^{C2D},\ \forall k \in \mathcal{K}, m \in \mathcal{M}$},
\STATE {$G^{D2D}_{k,k'}= \min_{d\in\mathcal{D}_k'} G_{k,k',d}^{D2D},\ \forall k \in \mathcal{K}, m \in \mathcal{M}$}

\STATE {$y_{k,m}=0, \ \forall k \in \mathcal{K}, m \in \mathcal{M} $}
\STATE {$P^{Cell}_m=P^{Cell}_{\max}, \ \forall m \in \mathcal{M} $}
\STATE {$P^{D2D}_{k,m}=0, \ \forall k \in \mathcal{K}, m \in \mathcal{M}$}
\STATE{$m=1$}
\WHILE {  $m \le M$}
\STATE {$\mathcal{K'} = \{ \forall k \in \mathcal{K}  |  \sum_{m=1}^M y_{k,m} < C_2\}$ }
\WHILE {$ \sum_{k=1}^K y_{k,m} < C_1$ or $\mathcal{K'} \ne \emptyset$ }
			\STATE{$k^*=\arg\min_{k\in \mathcal{K'} }\big(\sum_{k'=1}^K P^{D2D}_{k',m}G^{D2D}_{k,k'}+P^{Cell}_mG^{C2D}_{m,k}\big)$}
			\STATE{$y_{k^*,m}=1$}
		\STATE{Solve P3 to find $P^{D2D}_{k^*,m} $ and $P^{Cell}_m $ }
	\IF { P3 is feasible }
	 	\STATE {D2D  $k^*$ transmits on channel $m$}
		\STATE{$y_{k^*,m}=1$}
	\ELSE
		\STATE{$y_{k^*,m}=0$}
  	\ENDIF 
		 	\STATE {$\mathcal{K'} =\mathcal{K'} \setminus \{k^*\}$}	
\ENDWHILE
\STATE {$m=m+1$}
\ENDWHILE
\end{algorithmic}
\end{algorithm}

\subsection{Heuristic algorithm}
Since the complexity of the greedy algorithm is high, we propose a heuristic algorithm with less complexity in Algorithm~\ref{alg:hur}. In the following we explain some intuition behind the algorithm.
 
To increase cellular and D2D throughputs, it is desirable to have higher SINR. From (3) and (7), it can be deduced that having smaller values of $G_{k,m,d}^{C2D}$ and $G_{k,k',d}^{D2D}$ reduces interference from CU $m$ to D2D group $k$ and from D2D group $k$ to D2D group $k'$ respectively, and consequently results in higher $\beta^{D2D}_{k,m}$ and D2D throughput. Furthermore, higher values of $G^{Cell}_{m}$ lead to higher cellular throughput. 
Therefore, Algorithm~\ref{alg:hur}  tries to pair up a CU that has a high link gain to the BS and a D2D group that has low interference to the CU. 
Here, we assume that each CU sends the channel information between itself and D2D receivers through control channels to the BS. 

Starting from $m=1$, the outer loop in Algorithm~\ref{alg:hur} iterates through all CUs. For each $m$, the algorithm finds at most $C_1$ best D2D groups to share the channel $m$ in the inner loop.  Line 12 shows the criteria for choosing the D2D group that receives the minimum interferences from CU $m$ and all other D2D groups using the same channel. In line 14, based on the current value of $y_{k,m}$, problem P3 is solved to find the optimal transmission power for each CU and D2D group. If P3 is feasible, D2D group $k^*$ will reuse the channel $m$ and we have $y_{k^*,m}=1$, otherwise $y_{k^*,m}=0$ in line 20. In both cases, $k^*$ is removed from the D2D group list for the next iteration.
The inner loop stops iterating after finding $C_1$ D2D groups for CU $m$ or after at most $K$ iterations. It is worth mentioning that each D2D group cannot reuse more than $C_2$ CUs. That is accomplished by introducing $\mathcal{K'}$ that keeps track of all D2D groups with less than $C_2$ assigned channels in line 10.

In this algorithm, problem P3 is solved $M\times C_1$ times in the worst case, and thus the complexity of the heuristic algorithm is $\textbf {O}(M^2)+ \textbf {O}(M\times K\times f_3(| \mathcal{D_K}|))$. This is much less than the complexity of the greedy algorithm. However, as will be demonstrated in the simulation, the improvement in computation complexity comes at the cost of lower performance. 

We summarize the computational complexity of GBD, greedy and heuristic algorithms in Table~\ref{comp} in the worst case.

\begin{table}[h]
\centering
\caption{Worst case complexity comparison}
\begin{tabular}{|c|c|c|}
\rowcolor[HTML]{9B9B9B} 
\hline 
\textbf{Algorithm} & \textbf{Worst Case Complexity}  \\ 
\hline 
GBD & Exponential \\ 
\hline 
Greedy &  $\textbf {O}(\min\{M\times C_2, K\times C_1\}\times K \times M\times f_3(|\mathcal{D_K}|))$  \\ 
\hline 
Heursitic & $\textbf {O}(M^2)+ \textbf {O}(M\times K\times f_3(| \mathcal{D_K}|))$  \\ 
\hline 
\end{tabular} 
\label{comp}
\end{table}

\section {Performance Evaluation}
\label{sec:results}

We consider a single cell network as illustrated in Fig.~\ref{1}, where cellular users are uniformly distributed in the cell. We assume that the QoS requirements of all the CUs are satisfied before including D2D groups to the cell. 
The distance-based path loss and slow Rayleigh fading  are adopted as channel models. The proposed algorithms have been implemented in Matlab together with the CVX convex optimization package~\cite{cvx}.  Default parameters used in the simulations are given in Table~\ref{simP}. 
We run two sets of experiments to evaluate the performance of the proposed algorithms, namely, regularly placed D2D clusters and randomly placed D2D clusters.

\begin{table}[h]
\centering
\caption{Default Simulation Parameters}
\begin{tabular}{|l|l|}
\hline
\rowcolor[HTML]{9B9B9B} 
 \textbf{Parameter} &  \textbf{Value}\\ \hline
Cell radius (R)                            &1 km                             \\ \hline
Number of D2D receivers in each group           & 3                             \\ \hline
$P_{noise}$                               & -114 dBm                               \\ \hline
Pathloss exponent ($\alpha$)              & 3                                     \\ \hline
$P^{D2D}_{\max}$                          & 20 dBm                                \\ \hline
$P^{Cell}_{\max}$                         & 20 dBm                                \\ \hline
$\gamma_{th}$ =$\gamma^{Cell}_{th}$ =$\gamma^{D2D}_{th}$           & 10 dB                                 \\ \hline
D2D cluster size(r)                       & 50 m                               \\ \hline
\end{tabular}
\label{simP}
\end{table}

\begin{figure}[ ]
  \centering
    \includegraphics[width= 26em]{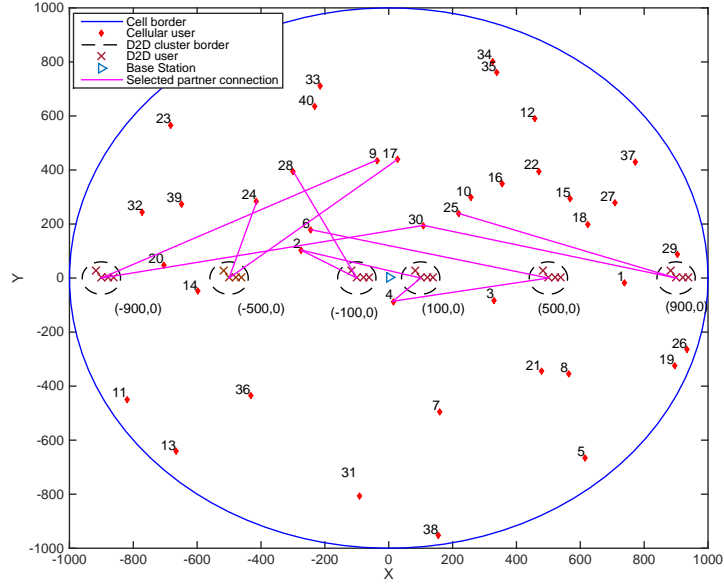}
      \caption{Regularly placed D2D clusters in a cell, $C_1=2, C_2=2$, $M=40$.} 
       \label{1}
\end{figure}

\paragraph {Regularly placed D2D clusters}In Fig.~\ref{1}, D2D groups are manually placed in six different locations and D2D transmitters and receivers are placed in the fixed locations within each group with radius $r$. This scenario allows us to have a better understanding of the channel selection for D2D users and how it is impacted by geographical spacing.  In the figure, D2D transmitters are labeled with their coordinates. 
The GBD algorithm finds the CU partner (or equivalent, the CU channel) for each D2D group among 40 CUs when $C_1=2$ and $C_2=2$. The straight lines in Fig.~\ref{1} connect D2D groups with their respective CU partners. As shown in the figure, the chosen CU partners, tend to be close to the base station to ensure the rate of the
CUs. Meanwhile, the CU partners are away from the respective D2D users to reduce mutual interference between the CUs and the D2D users. Note that even for CUs at the cell edges, their SINR constraints are satisfied as guaranteed by P1. 

Fig.~\ref{1.1} compares the maximum cellular throughput (without D2D users), $R^{Cell}_{\max}$, the throughput of cellular users (with D2D users), $R^{Cell}$, and  D2D throughput, $R^{D2D}$, defined as follows,

 \begin{equation}
 \label {eq:66}
R^{Cell}_{\max}=\sum_{m=1}^M\log_2\left(\frac{P^{Cell}_{\max} G^{Cell}_{m}} {P_{noise}} \right),
\end{equation}

 \begin{equation}
 \label {eq:66}
R^{Cell}=\sum_{m=1}^MR^{Cell}_m,
\end{equation}
 \begin{equation}
 \label {eq:66}
R^{D2D}=\sum_{k\in \mathcal{A}} R^{D2D}_k,
\end{equation}
%
where $\mathcal{A}$ is the set of D2D groups that are allowed to reuse at least cellular channel. 
As can be observed in Fig.~\ref{1.1}, the overall network throughput, $R^{sum}=R^{Cell}+R^{D2D}$, is greater than the maximum throughput before including D2D users, $R^{Cell}_{\max}$. With the introduction of D2D users, the overall throughput increases by 25\% to 125\%. This comes at the cost of reduced cellular throughput as $R_{max}^{Cell} > R^{Cell}$ since adding D2D users causes interference to cellular users and decreases their throughput. However, the reduction is relatively small, compared to the D2D throughput.
Moreover, although a larger D2D cluster size leads to lower D2D channel gain and lower D2D throughput, it does not affect the cellular throughput very much.

\begin{figure}[ ]
  \centering
    \includegraphics[width= 25em]{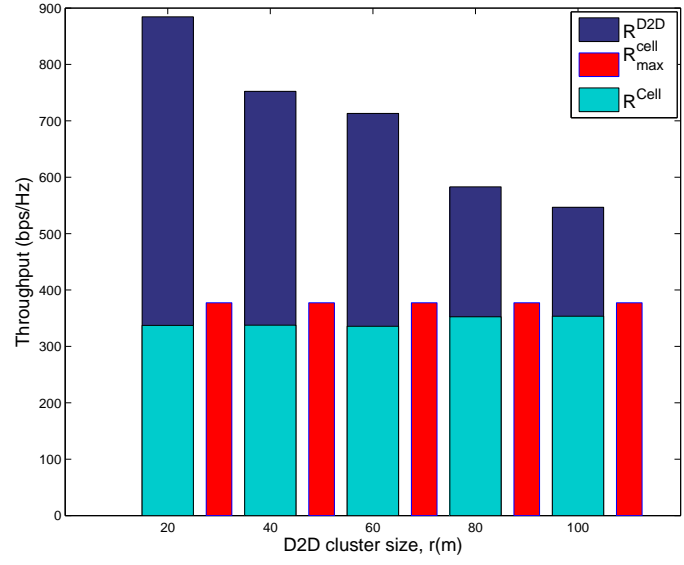}
      \caption{Throughput comparison for different cluster sizes, $C_1=2, C_2=2$, $M=40$.}
       \label{1.1}
\end{figure}

\begin{figure}[]
  \centering
    \includegraphics[width= 25em]{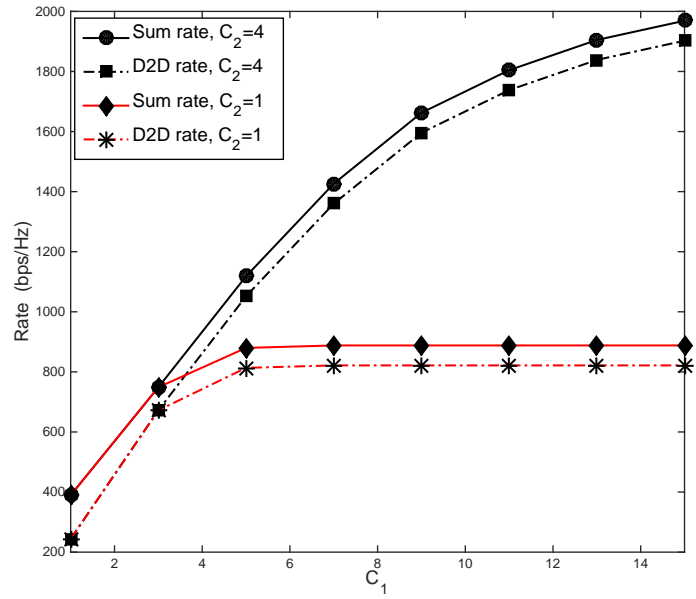}
      \caption{Throughput comparison for different values of $C_1$ and $C_2$, $M=20$. }
       \label{1.2}
\end{figure}
Fig.~\ref{1.2} shows D2D and sum rates versus $C_1$ for different values of $C_2$. Both rates increase with $C_1$ since  the number of available channels for each D2D group increases and hence D2D rate increases. However, when $C_2=1$, both the D2D and sum rates flatten out after a certain value of $C_1$. In this case, each CU can serve at most one D2D group, and increasing $C_1$ does not increase the rate since there are not enough channels to allow all the D2D groups to reuse $C_1$ channels.
Also, from this figure we see that cellular throughput, which is the difference between the sum rate and the D2D rate, decreases as $C_1$ increases. This is because of the fact that the interference from D2D groups on CUs increases with $C_1$.
On the other hand, increasing $C_2$ increases the D2D and sum rate for higher values of $C_1$ since each CU can serve more D2D groups and hence there are more available channels for D2D groups.
However, for lower values of $C_1$, since there are enough CUs in the cell to be reused by D2D groups, increasing $C_2$ does not change the D2D and sum rates significantly.

\paragraph {Randomly placed D2D users}In the second set of experiments, we follow the clustered distribution model in~\cite{cluster}, where clusters of radius $r$ are randomly located in a cell and the D2D users in each group are randomly distributed in the corresponding cluster.  Four metrics are used to evaluate the performance: 
the sum throughput, $R^{sum}$, the D2D throughput, $R^{D2D}$, the success rate, and the fairness index.
The success rate is defined as the ratio of the number of D2D groups that found their CU partners ($|\mathcal{A}|$) and the total number of D2D groups. Fairness index  is defined as follows,

 \begin{equation}
 \label {eq:55}
f(R^{D2D}_1,R^{D2D}_2,\dots,R^{D2D}_k )=\frac{(\sum_{k\in\mathcal{A} }R^{D2D}_k)^2}{|\mathcal{A}|\sum_{k\in\mathcal{A} }{(R^{D2D}_k})^2}
\end{equation}
The fairness index is a positive number with the maximum value of 1 suggesting an equal D2D throughput among all feasible D2D groups.

The results in this section have been generated for two sets of $C_1$ and $C_2$ values: in part (a) of all the figures, $C_1=4$ and $C_2=3$; and in part (b), $C_1=1$ and $C_2=1$. In the case of $C_1=1$ and $C_2=1$, both GBD and the matching-based algorithm return the same results since both are optimal.
In our previous work,~\cite{Meshgi15}, we have adapted the heuristic scheme in~\cite{M.zul} for multicast D2D and compared it against proposed scheme when $C_1 = 1$ and $C_2 = 1$. Numerical results in~\cite{Meshgi15} show that our proposed heuristic outperforms the resource allocation algorithm in~\cite{M.zul}, and thus evaluation of the heuristic in~\cite{M.zul} is omitted here. 


\begin{figure}[htp]
  \centering
  \subfigure[$C_1=4, C_2=3$]{\includegraphics[width= 20em]{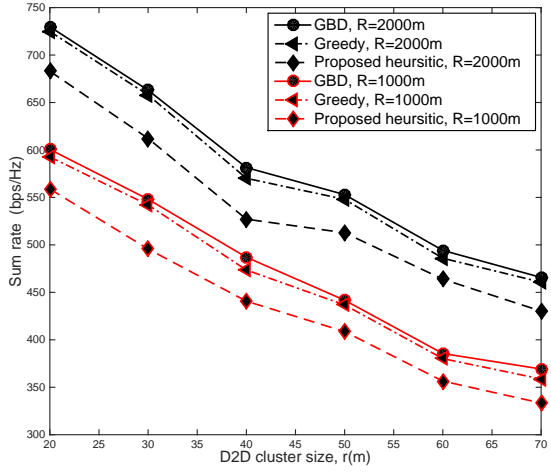}}\label{f2.1}
  \subfigure[$C_1=1, C_2=1$]{\includegraphics[width= 20em]{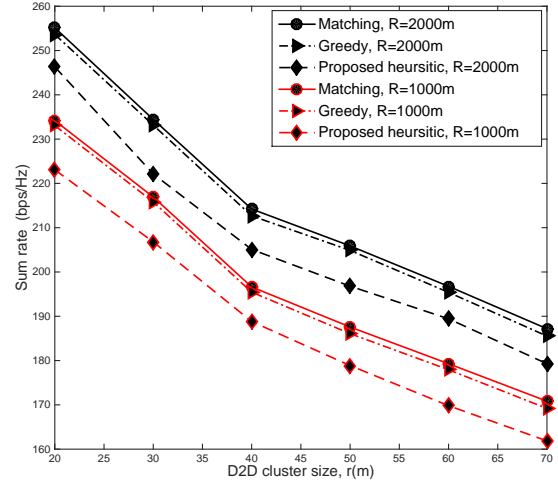}}\label{f2.2}
  \caption{Average sum throughput versus D2D cluster radius for different cell radii ($R$), $M=10, K=4$ }
  \label{f2}
\end{figure}

\begin{figure}[htp]
  \centering
  \subfigure[$C_1=4, C_2=3$]{\includegraphics[width= 20em]{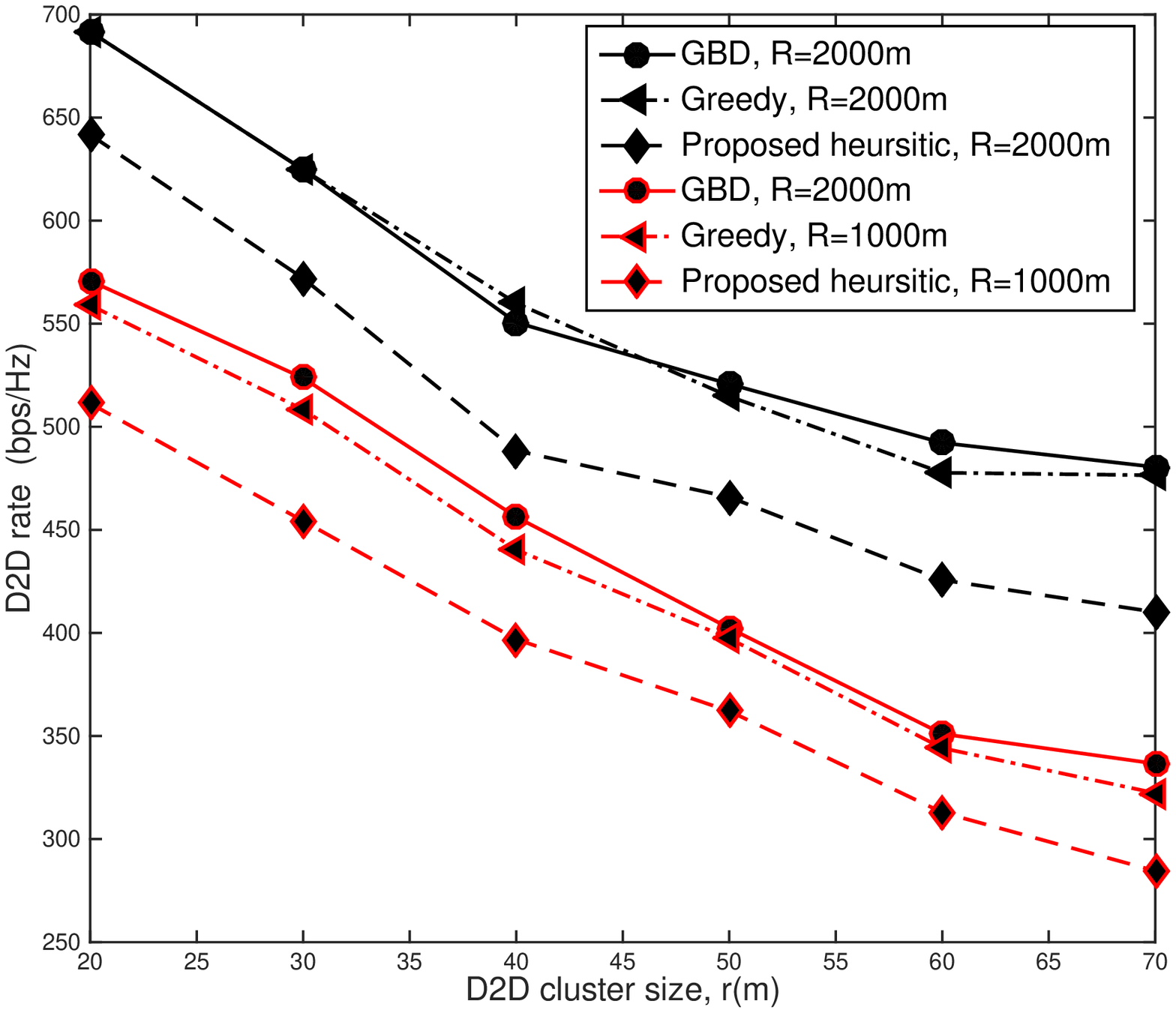}}\label{f3.1}
  \subfigure[$C_1=1, C_2=1$]{\includegraphics[width= 20em]{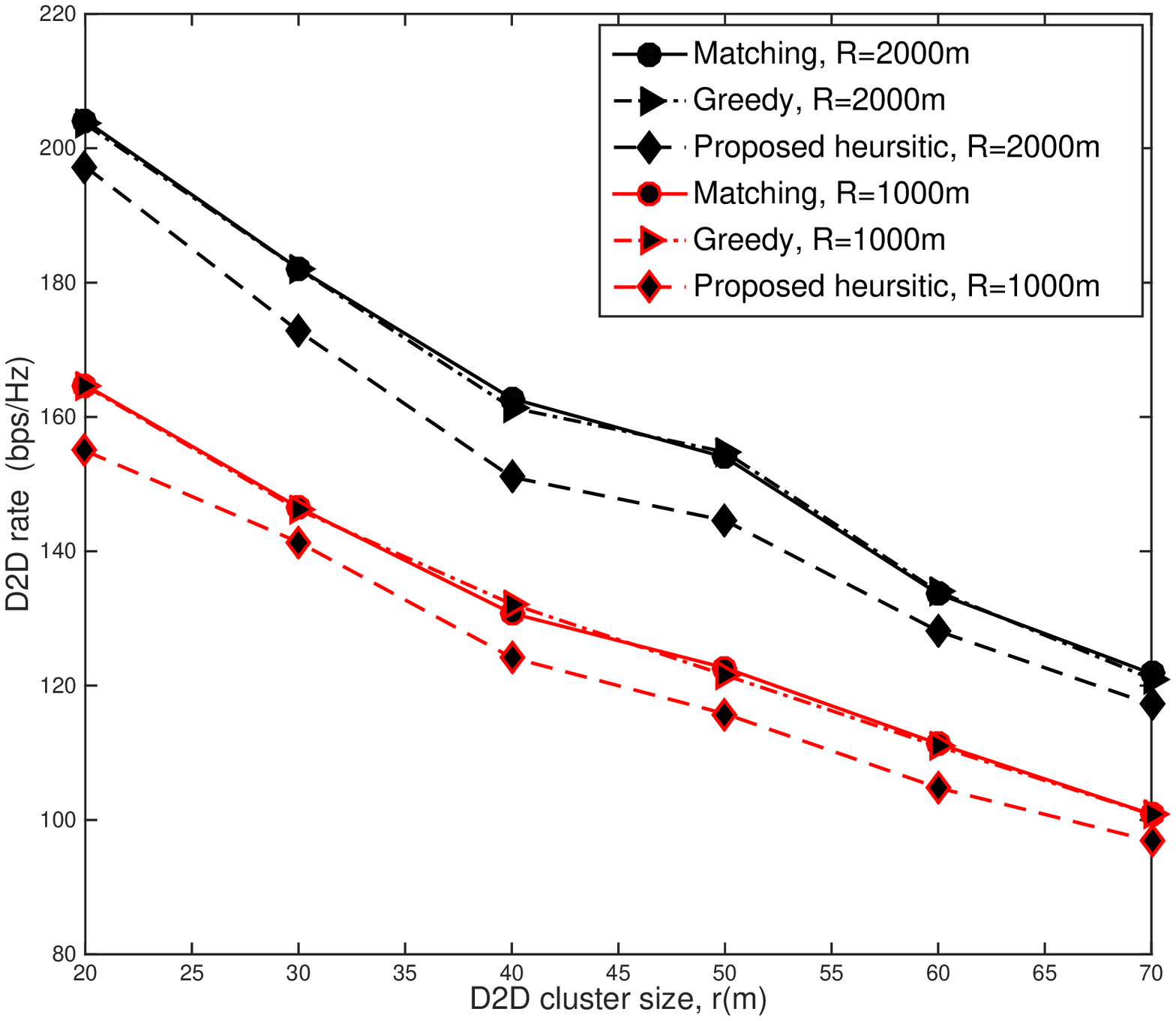}}\label{f3.2}
  \caption{Average D2D throughput versus D2D cluster radius for different cell radii ($R$), $M=10, K=4$}
  \label{f3}
\end{figure}

\begin{figure}[htp]
  \centering
  \subfigure[$C_1=4, C_2=3$]{\includegraphics[width= 20em]{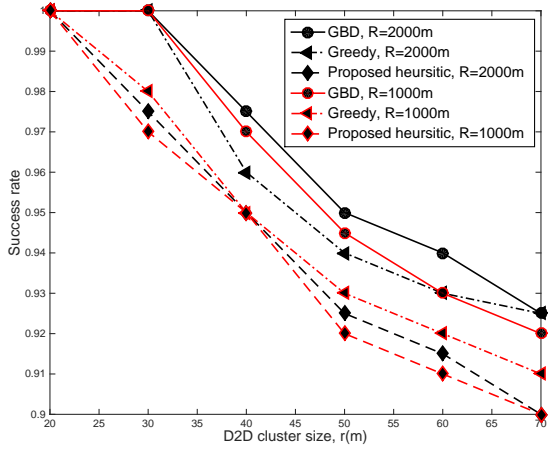}}\label{f4.1}
  \subfigure[$C_1=1, C_2=1$]{\includegraphics[width= 20em]{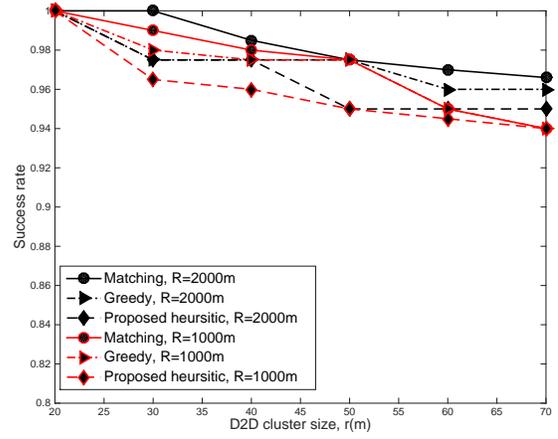}}\label{f4.2}
  \caption{Average D2D success rate versus D2D cluster radius for different cell radii ($R$), $M=10, K=4$}
  \label{f4}
\end{figure}

\begin{figure}[htp]
  \centering
  \subfigure[$C_1=4, C_2=3$]{\includegraphics[width= 20em]{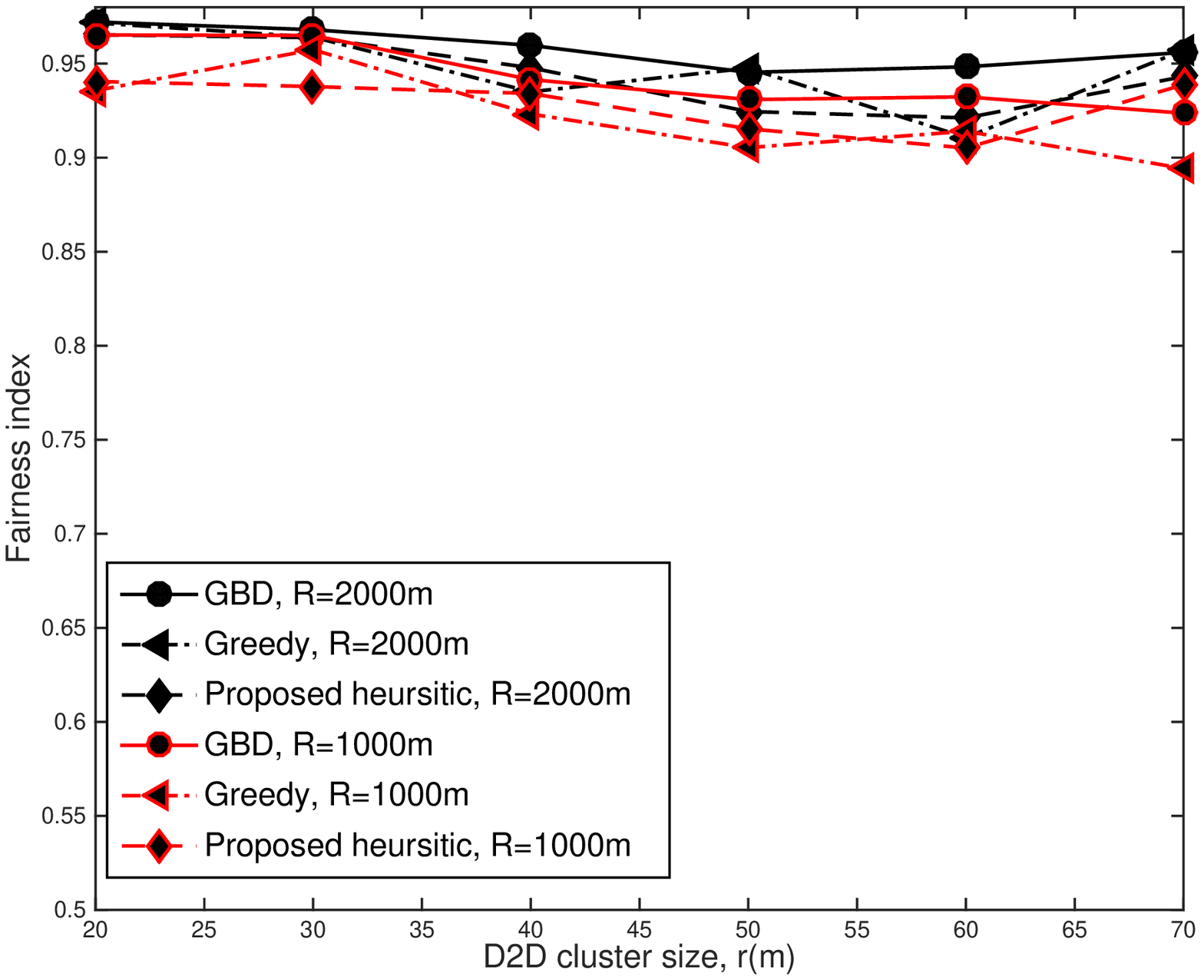}}\label{f5.1}
  \subfigure[$C_1=1, C_2=1$]{\includegraphics[width= 20em]{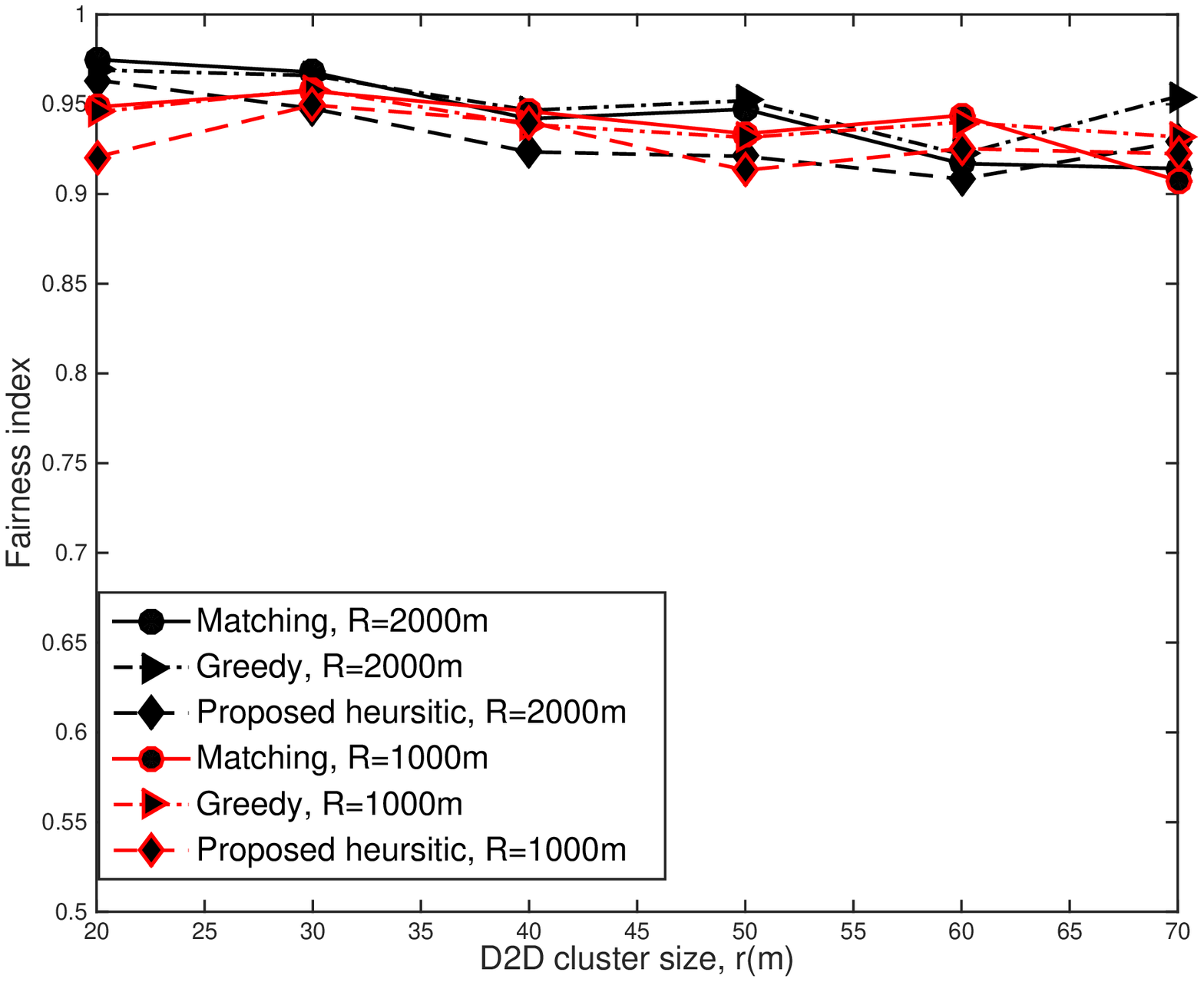}}\label{f5.2}
  \caption{Average fairness index versus D2D cluster radius for different cell radii ($R$), $M=10, K=4$}
  \label{f5}
\end{figure}

Figs.~\ref{f2} --~\ref{f5} compare the performance of GBD, the greedy and the heuristic algorithms for different D2D cluster sizes ($r$) and different cell radii ($R$). From these figures, we observe that both the sum and the D2D throughput as well as the success rate decrease
with the D2D cluster size. Since the channel gain of D2D link decreases when the cluster radius increases, more transmission power is required for the D2D groups to satisfy the SINR threshold constraint. This in turn causes more interference to the reused CU partner. Furthermore, it is seen from these figures that the sum throughput, the D2D throughput and the success rate of all three algorithms increase with the cell radius. This is because increasing the cell radius increases the distance between the CUs and D2D receivers and also the average distance of individual nodes to the BS. Hence, the interference from CUs to D2D receivers and the interference from D2D transmitters at the BS is decreased. Recall that the D2D rate is the maximum throughput achieved by the admitted D2D groups. It is worth mentioning that increasing the cell size leads to reduction in the cellular throughput due to the decreased link gain between the CUs and the base station. However, with the current simulation parameters, $R^{D2D}$ is the dominating part in the sum rate and therefore $R^{sum}$ increases with the cell size in both parts (a) and (b). 

It can be  also seen from Fig.~\ref{f2} that the optimal solutions, GBD algorithm for part (a) and matching-based algorithm for part (b), has the highest sum rates. In comparison, the greedy algorithm achieves close-to-optimal sum rate, while the heuristic algorithm has a lower sum rate compared to the other two algorithms but it has the lowest complexity among them.
Note that in Fig. ~\ref{f3}, the D2D rate of the greedy algorithm exceeds that of the optimal solution for some D2D cluster sizes. This does not contradict the optimality of GBD since the objective of P1 is to maximize the sum rate not the D2D rate.


Fig.~\ref{f5}  shows that the D2D fairness indices achieved by all algorithms are greater than 90\%. Note that the fairness index calculates the fairness among all \textit{admitted} D2D groups. Therefore, we can conclude that there is not much difference among  D2D rates of all admitted D2D groups.


In Figs.~\ref{6} --~\ref{9} the performance of all proposed algorithms
for different SINR thresholds ($\gamma^{D2D}_{th}=\gamma^{Cell}_{th}=\gamma_{th}$) with different numbers of CUs ($M$) is shown. It is seen that increasing the SINR threshold leads to decreasing sum rates, D2D rates, and success rates since it limits the chances for D2D groups to find CU partners.  It can be also observed that the total D2D throughput  improves slightly with increasing number of CUs since there are more potential candidates for D2D groups to reuse.
%
%


%


\begin{figure}[htp]
  \centering
  \subfigure[$C_1=4, C_2=3$]{\includegraphics[width= 20em]{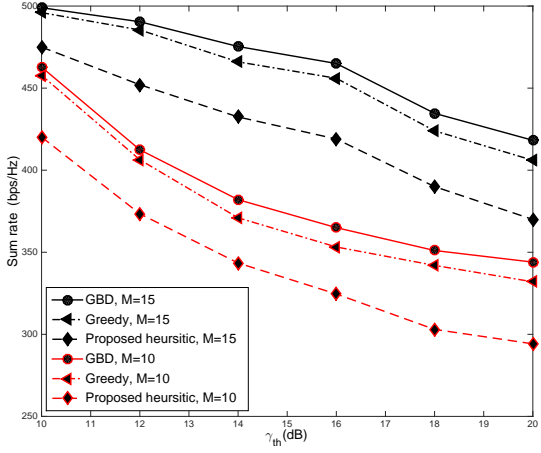}}\label{6.1}
  \subfigure[$C_1=1, C_2=1$]{\includegraphics[width= 20em]{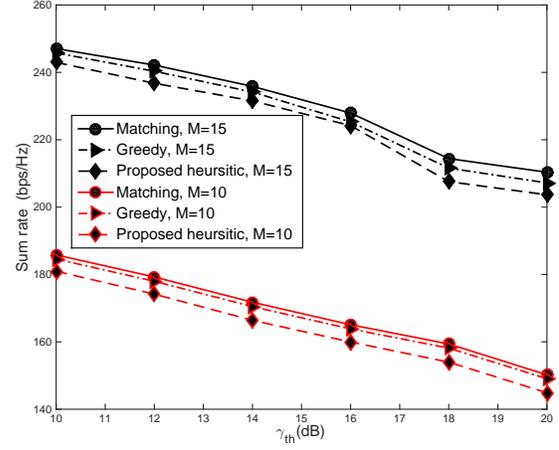}}\label{6.2}
  \caption{Average sum throughput versus $\gamma_{th}$ for different number of cellular users ($M$), $R=1000m, K=4$}
  \label{6}
\end{figure}

\begin{figure}[htp]
  \centering
  \subfigure[$C_1=4, C_2=3$]{\includegraphics[width= 20em]{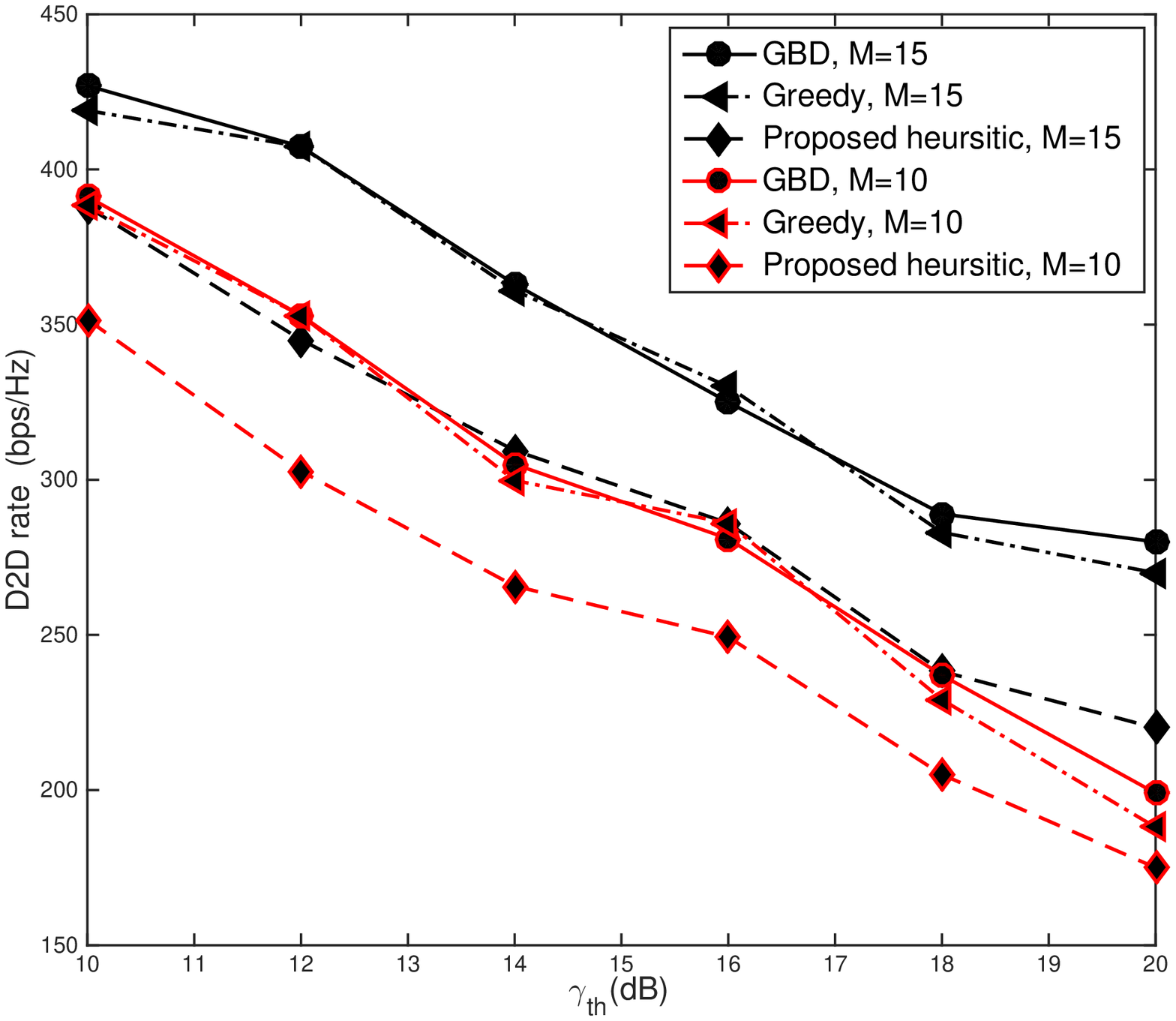}}\label{7.1}
  \subfigure[$C_1=1, C_2=1$]{\includegraphics[width= 20em]{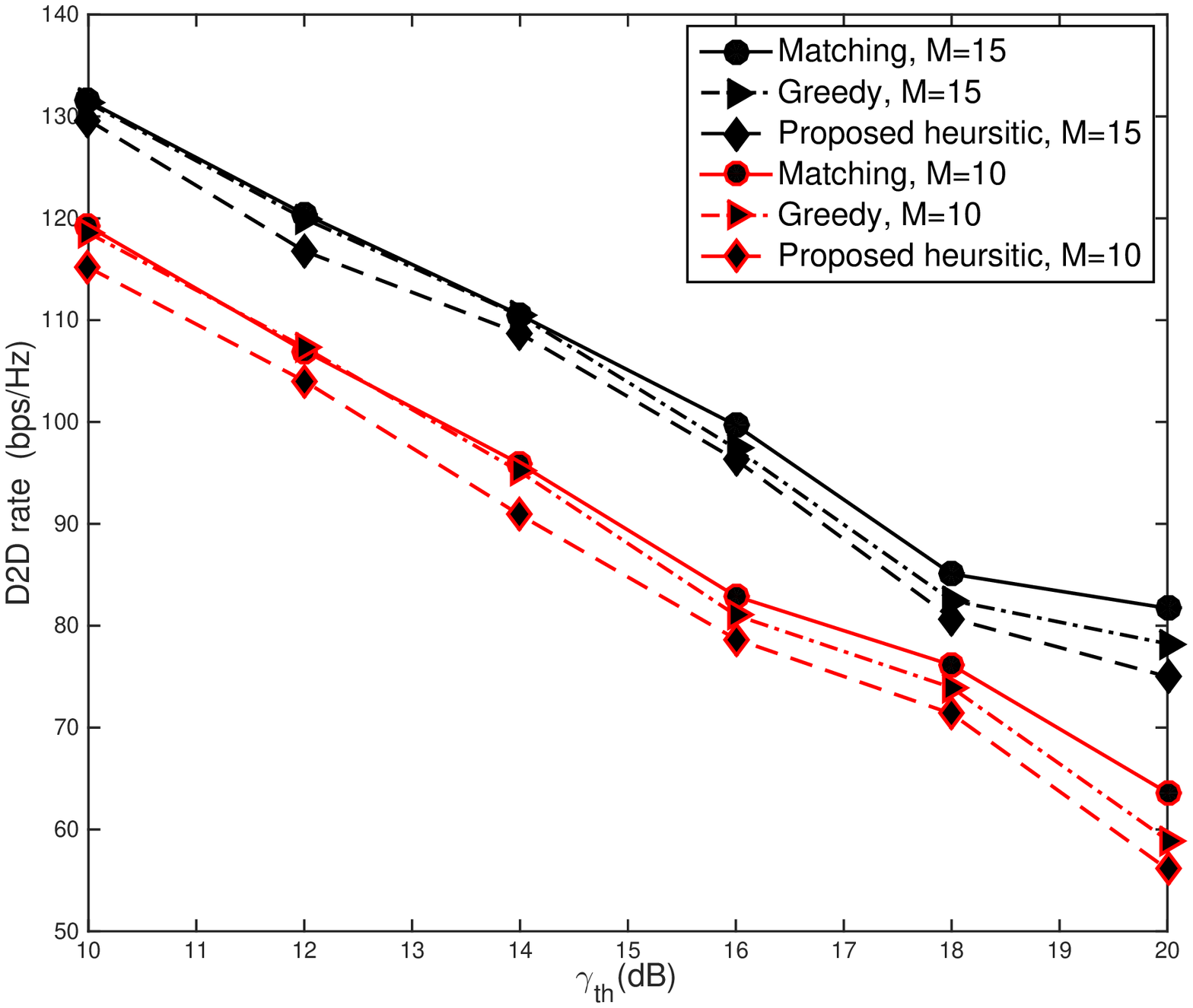}}\label{7.2}
  \caption{Average D2D throughput versus $\gamma_{th}$ for different number of cellular users ($M$), $R=1000m, K=4$}
  \label{7}
\end{figure}

\begin{figure}[htp]
  \centering
  \subfigure[$C_1=4, C_2=3$]{\includegraphics[width= 20em]{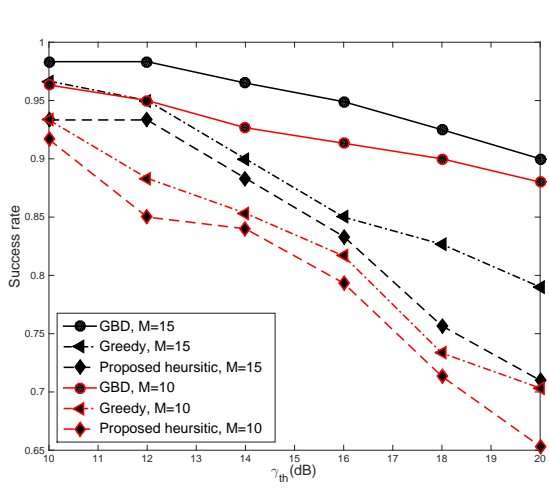}}\label{8.1}
  \subfigure[$C_1=1, C_2=1$]{\includegraphics[width= 20em]{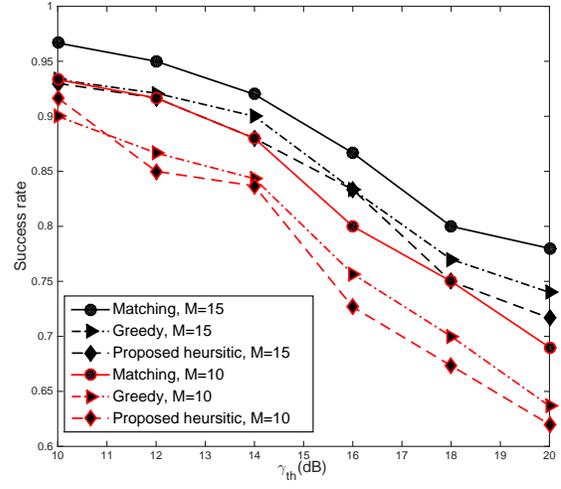}}\label{8.2}
  \caption{Average D2D success rate versus $\gamma_{th}$ for different number of cellular users ($M$), $R=1000m, K=4$}
  \label{8}
\end{figure}

\begin{figure}[htp]
  \centering
  \subfigure[$C_1=4, C_2=3$]{\includegraphics[width= 20em]{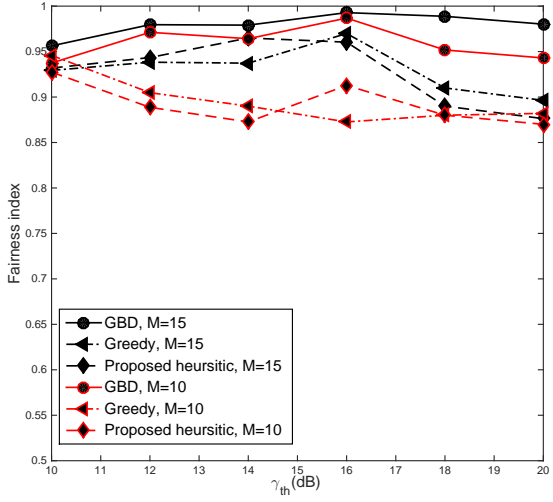}}\label{9.1}
  \subfigure[$C_1=1, C_2=1$]{\includegraphics[width= 20em]{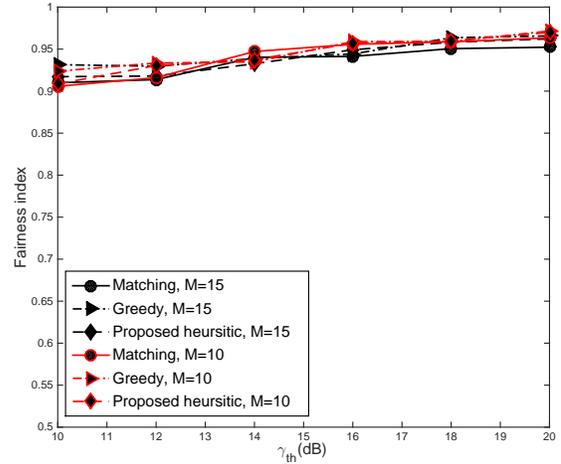}}\label{9.2}
  \caption{Average fairness index versus $\gamma_{th}$ for different number of cellular users ($M$), $R=1000m, K=4$}
  \label{9}
\end{figure}

\section{Conclusions}
\label{sec:conclusions} 
In this paper, we  considered joint power and channel allocation for multicast D2D communications sharing uplink resource in a fully loaded cellular network. To maximize the overall throughput while guaranteeing the QoS requirements of both CUs and D2D groups, we formulated the optimization problem and found the optimal solution using GBD. Then, we solved a special case when each D2D group can reuse the channels of at most one CU and each CU can share their channels with at most one D2D group, using maximum weight bipartite matching algorithm. Finally, a greedy algorithm and a low-complexity heuristic algorithm were also proposed. We performed extensive simulations with different parameters such as SINR threshold, cell size, D2D cluster size, and number of CUs. Results showed that the greedy algorithm has close-to-optimal performance. In comparison, our proposed heuristic algorithm has good performance (but worse than that of the greedy) with lower computation complexity. 


\bibliographystyle{IEEEtran}
\bibliography{references}        

\end{document}